# Thickness-dependent Dielectric Constant of Few-layer In$_2$Se$_3$ Nano-flakes

Di Wu[1,*], Alexander J. Pak[2,*], Yingnan Liu[1], Yu Zhou[3], Xiaoyu Wu[1], Yihan Zhu[4], Min Lin[3], Yu Han[4], Yuan Ren[1], Hailin Peng[3], Yu-Hao Tsai[2], Gyeong S. Hwang[2,**], Keji Lai[1,**]

[1]Department of Physics, University of Texas at Austin, Austin TX 78712

[2]Department of Chemical Engineering, University of Texas at Austin, Austin TX 78712

[3]College of Chemistry and Molecular Engineering, Peking University, Beijing 100871, China

[4]Physical Sciences and Engineering Division, King Abdullah University of Science and Technology, Thuwal 23955-6900, Saudi Arabia

**ABSTRACT:** The dielectric constant or relative permittivity ($\varepsilon_r$) of a dielectric material, which describes how the net electric field in the medium is reduced with respect to the external field, is a parameter of critical importance for charging and screening in electronic devices. Such a fundamental material property is intimately related to not only the polarizability of individual atoms, but also the specific atomic arrangement in the crystal lattice. In this letter, we present both experimental and theoretical investigations on the dielectric constant of few-layer In$_2$Se$_3$ nano-flakes grown on mica substrates by van der Waals epitaxy. A nondestructive microwave impedance microscope is employed to simultaneously quantify the number of layers and local electrical properties. The measured $\varepsilon_r$ increases monotonically as a function of the thickness and saturates to the bulk value at around 6 ~ 8 quintuple layers. The same trend of layer-dependent dielectric constant is also revealed by first-principle calculations. Our results of the dielectric response, being ubiquitously applicable to layered 2D semiconductors, are expected to be significant for this vibrant research field.

The rapid rise of graphene in the past decade has led to an active research field on two-dimensional (2D) layered materials.[1,2] Of particular interest here are layered semiconductors, such as many metal chalcogenides, for their roles as gate dielectrics or channel materials in next-generation electronics.[3] Due to the strong intralayer covalent bonding and weak van der Waals (vdW) interactions, most physical properties are already anisotropic in the bulk form, with a clear 3D-2D crossover when approaching the monolayer thickness. In particular, the number of layers ($n$) in a thin-film 2D system is expected to strongly influence its dielectric constant, a fundamental electrical property that determines the capacitance and charge screening in electronic devices.[4-6]

The 2D material in this study is the layered semiconducting chalcogenide $In_2Se_3$, a technologically important system for phase-change memory, thermoelectric, and photoelectric applications[7]. The In-Se phase diagram is among the most complex ones in binary compounds. Even at the exact stoichiometry of In:Se = 2:3, multiple phases can occur under different temperatures and pressures.[8-11] By controlling the synthesis parameters or thermal/electrical pre-treatment processes, several phases (superlattice, simple hexagonal α-phase, simple hexagonal β-phase, and amorphous state) with vastly different electrical conductivity can coexist at the ambient condition,[12-14] which explains the research interest of $In_2Se_3$ as a prototypical phase-change material.[7,12-18] In addition, the lattice constant of $In_2Se_3$ matches well with $Bi_2Se_3$, which is heavily investigated for its high thermoelectric figure-of-merit and topological insulator nature.[19,20] The chemical and structural compatibility between the two chalcogenides enables the growth of $In_2Se_3/Bi_2Se_3$ heterostructures.[21,22] For all aforementioned applications, the dielectric constant of few-layer $In_2Se_3$ is an important physical parameter that must be fully characterized as a function of $n$. We further emphasize that, although $In_2Se_3$ is used as a model system in this work, neither the experimental technique nor the theoretical analysis is limited to this particular system and the conclusions presented here are generally applicable for the fast growing field of 2D vdW materials.

The few-layer $In_2Se_3$ nano-flakes were grown by van der Waals epitaxy on mica substrates,[23] as schematically shown in Fig. 1(a). Detailed growth parameters and characterizations of the structural and electrical properties can be found in Ref. [23]. When the sample was rapidly cooled down (> 100°C/min) from the deposition temperature, a superlattice phase can be observed by transmission electron microscopy (TEM),[12,13,23] which exhibits metallic behaviors with a high conductivity in the order of $10^5$ S/m. At much slower cooling rates (< 5°C/min), on the other hand, the simple hexagonal lattice was obtained, showing the usual semiconducting behavior with a room-temperature conductivity of $10^{-2} \sim 10^{-3}$ S/m.[23] In this work, we have carefully controlled the growth conditions and screened the samples such that only the semiconducting α-phase $In_2Se_3$ was studied. No thermal pretreatment has been applied to the as-grown samples so neither β-phase nor amorphous phase was involved. Different batches of samples were measured, with no discernible difference with respect to each other.

Fig. 1(b) shows a typical scanning electron microscopy (SEM) image of as-grown $In_2Se_3$ flakes on mica. Most flakes are triangular or hexagonal in shape, while some thin pieces ($n \leq 3$) are rounded at the corners. High resolution transmission electron microscopy (HRTEM) was carried out to image the [001] in-plane lattice fringes and [100], [120] cross-sections, which were obtained by cutting the samples along and perpendicular to one side of a triangle (dashed lines in Fig. 1b), respectively. As shown in Fig. 1(c) and Supporting Information Fig. S1，the crystal symmetry along the [001] projection is P3m1, while the P1 and Pm symmetries along [100] and [120] directions give rise to specific electron diffraction or FFT patterns, suggesting that $In_2Se_3$ has the R3m structure. Although multiple phases were observed in previous work,[13, 19] extensive TEM images reveal that the as-grown samples in this study are predominantly α-phase with R3m space group, which is also confirmed by the Raman spectra in Fig. 1(d). Two vibrational modes, E and $A_1$, can be observed at around 144 $cm^{-1}$ and 237 $cm^{-1}$.[24] Similar to $MoS_2$,[25] the E mode shows a red shift and the $A_1$ mode exhibits a blue shift as the thickness increases from 3 nm to 13 nm. The experimental details are further described in the method section.

A microwave impedance microscope (MIM) based on the atomic-force microscopy (AFM) platform[13, 26-28] was employed to measure the thickness and dielectric response of few-layer $In_2Se_3$ flakes, as illustrated in Fig. 2(a). During the contact-mode AFM scans, an excitation signal at 1 GHz is delivered to the tip and the reflected wave is detected by the microwave electronics to form the imaginary (MIM-Im) and real (MIM-Re) components of the MIM output. Because the relevant tip and sample dimensions are much smaller than the free-space wavelength ($\lambda$ = 30 cm) at 1 GHz, the extreme near-field interaction can be modeled as a lumped-element circuit, from which the local dielectric constant and conductivity of the sample can be deduced.[27]

Figs. 2(c) and 2(d) show the simultaneously taken topography and MIM images of several $In_2Se_3$ nano-flakes. Aside from some insulating particles (bright in AFM and dark in MIM-Im), both the In$_2$Se$_3$ flakes and the mica substrate are atomically flat, confirming the layered vdW characteristics. Since each Se-In-Se-In-Se quintuple layer is about 1 nm thick[23], which was also confirmed by the AFM measurements (Supporting information Fig. S2), the different terraces of the sample in Fig. 2(c) can be designated as $n$ from 3 to 6. Interestingly, the corresponding MIM-Im image shows distinct contrast over the mica substrate as a function of the flake thicknesses. For $n = 2$ or 3, the MIM signal on the flake is lower than that on the substrate, indicating that the dielectric constant of ultra-thin In$_2$Se$_3$ is smaller than $\varepsilon_{r, mica} \approx 6$.[29] On the other hand, the regions of the flake with $n = 4$ are hardly seen over the background, while the 5-layer section is clearly visible. For thicker flakes with $n \geq 6$, the MIM-Im signals are well above that of the mica substrate, consistent with the relatively large bulk In$_2$Se$_3$ dielectric constant of $\varepsilon_r = 17$.[30] The same trend was observed in other flakes, as enumerated in Fig. 2(d). The complete set of AFM and MIM images can be found in Supporting Information Fig. S3. For all nano-flakes we measured, no signal above the noise level was detected in the MIM-Re channel, suggesting that the MIM signal is purely due to permittivity contrast over the mica substrate, rather than the negligible conductivity in the as-grown $\alpha$-phase In$_2$Se$_3$.[23]

Quantitative analysis of the AFM/MIM data is shown in Fig. 3. From the linear fit of the AFM data (Figure S2), the thickness of a single In$_2$Se$_3$ quintuple layer is about 1.02 nm. The MIM-Im signals, however, show a clear downward kink at around 6 layers when approaching the monolayer limit. To understand the results, we performed numerical simulations using a finite-element analysis (FEA) software COMSOL4.4[26] and the details are included in Supporting Information Fig. S4. Because of the axisymmetric quasi-static electric field around the tip apex, we can only obtain the effective isotropic $\varepsilon_{r,eff}$, rather than separately determine the in-plane and out-of-plane dielectric constants in In$_2$Se$_3$. As demonstrated in Fig. 3(b), within the experimental errors, the effective dielectric constant increases monotonically from 2 to 6 layers and saturates to the bulk value after $n = 8$. We emphasize that, while a similar thickness-dependent dielectric constant in MoS$_2$ has been recently reported by parallel-plate experiments,[31] the MIM approach is noninvasive to the materials so that as-grown samples can be measured without patterning any contact electrodes.

To further understand the thickness-dependence of the macroscopic $\varepsilon_r$, we carried out first-principles calculations using density functional theory (DFT) with the Heyd-Scuseria-Ernzerhof hybrid functional (HSE06).[32] Here, we only considered free-standing QLs of In$_2$Se$_3$ as a model system and calculated the electronic contribution to permittivity (*i.e.*, $\varepsilon^\infty$ or the so-called ion-clamped dielectric constant) in order to understand the role of interlayer interactions on $\varepsilon_r$. Note that $\varepsilon^\infty$ underestimates the static $\varepsilon_r$ by the portion of the ionic contribution but is sufficient to reveal important trends in the thickness-dependent behavior. Each QL slab was modeled in the R3m space group, containing In atoms in both tetrahedral and octahedral sites[8,31,33]; note that the QLs are stacked in the AB fashion, as described in the Methods section and depicted in Fig. 4(b). Figure 4(a) shows the calculated variation of the in-plane $\varepsilon_{xy}^\infty$ ($\varepsilon_x^\infty = \varepsilon_y^\infty$) and out-of-plane $\varepsilon_z^\infty$ as $n$ increases, indicating that $\varepsilon^\infty$ monotonically increases as $In_2Se_3$ transitions from the monolayer to the bulk crystal. Such thickness dependent behavior is consistent with that predicted for transition metal dichalcogenides.[5,34,35] This result suggests that the intrinsic polarizability within the QLs tends to increase with $n$. In addition, we find that $\varepsilon^\infty$ exhibits clear anisotropy that is lower in the direction perpendicular to the QL surface (*i.e.*, *c*-axis), which is consistent with other 2D materials.[5, 34, 35]

The Born effective (dynamical) charge tensor ($Z_\alpha^*$) along principal directions $\alpha$ was calculated to further analyze the local dipole moment behavior. $Z_\alpha^*$ is a measure of the macroscopic current and spontaneous polarization that result from the displacement of ions.[36] A summary of the computed in-plane $Z_{xy}^*$ and cross-plane $Z_z^*$ for the bulk and $n = 2$ cases can be found in Table 1, following the naming convention from Fig. 4(b). First, we note that $Z_{xy}^*$ tends to be close (within 35%) to the nominal charges of Se (-2) and In (+3), indicating limited in-plane charge transfer in response to an external field. In addition, $Z_{xy}^*$ remains effectively the same in the $n = 2$ and the bulk cases. However, a clear anisotropy is present in which $|Z_z^*| < |Z_{xy}^*|$. In fact, the magnitude of $Z_z^*$ exhibits a large suppression with respect to the nominal charges (as much as 85%) that is increasingly repressed as $n$ decreases, suggesting the induction of a local dipole in the out-of-plane direction that is consistent with the anisotropic $\varepsilon^\infty$ shown in Fig. 4(a). Interestingly, this behavior is fundamentally different from 3D crystalline materials, such as perovskites that typically exhibit anomalously large $Z^*$; in these cases, the large $Z^*$ is attributed to the covalence between anion and cation atoms.[36-38] Instead, our results suggest that the suppression in $Z_z^*$ is due to the layered nature of $In_2Se_3$, specifically the weakness of the interlayer Se-Se interaction with respect to the intralayer In-Se bond. As such, $Z_z^*$ approaches the nominal charges as $n$ increases (*i.e.*, increasing polarizability) and the $Z_z^*$ of the inner Se ($Se_3$) tends to be largest within each layer as the Se atom is fully coordinated.

This analysis demonstrates the sensitivity of the dielectric properties to the interlayer interactions. On the other hand, we note that the experimental data show a more abrupt increase of $\varepsilon_r$ around $n = 4 \sim 6$ than that predicted by the theoretical calculations. One factor we did not include in the model systems is the possible interaction between the QLs and the mica substrate, which may strongly influence $\varepsilon_r$ when $n$ is small according to our hypothesis. Therefore, it may be insightful to explore $In_2Se_3$ on different substrates to further understand the role of the substrate-QL interaction with respect to QL-QL interactions on its optical properties.

In summary, we have observed a layer-dependent dielectric constant of few-layer $In_2Se_3$ nano-flakes via MIM technique. The dielectric constant of $In_2Se_3$ rises monotonically as the layer number increases from 2 to 6 and approaches the bulk value beyond $n = 6$. First-principle calculations suggest that the smaller dielectric constant of few-layer flakes results from the suppression of cross-plane polarization due to the weak interlayer Se-Se interaction relative to the inner In-Se interaction, which is a specific feature of layered materials. Our finding is not only significant for fundamental research, but also beneficial for multifunctional nano-electronics based on the tunable dielectric constant of 2D materials.

**METHODS**

**Growth of $In_2Se_3$:** The van der Waals epitaxy growth of semiconducting $In_2Se_3$ flakes on fluorophlogopite mica were carried out in a horizontal tube furnace (Lindberg Blue M HTF55667C) equipped with a 1-inch-diameter quartz tube. The $In_2Se_3$ powder source (99.99%, Alfa Aesar) was placed at the hot center of the tube furnace heated to 690-750 ºC. The vapor was carried downstream by 30-200 sccm Ar gas, and $In_2Se_3$ flakes were deposited on the substrate 7~12 cm away from heating center.

**Raman Spectrum Measurements:** Raman spectroscopy was carried out using Witec Alpha 300 micro-Raman confocal microscope and a laser operating at wavelength of 488 nm. To avoid the Raman peaks of mica substrate, we transferred the $In_2Se_3$ from mica to $SiO_2$ (300 nm)/Si substrate using the method which transfers graphene and other 2D materials.[23]

**Microwave Impedance Microscopy measurements:** The MIM in this work is based on an AFM platform (ParkAFM XE-70). The customized shielded cantilevers are commercially available from PrimeNano Inc.[28] Finite-element analysis was performed using the commercial software COMSOL4.4.

**Density functional theory calculations:** The ground-state geometries and static dielectric constants were calculated using DFT within the HSE06 generalized gradient approximation,[32, 39] as implemented in the Vienna *Ab-initio* Simulation Package (VASP).[40] The projector augmented wave method[41, 42] was employed to describe the interaction between the core and valence electrons using a plane-wave basis set with a cutoff of 400 eV. Dispersion corrections were included using the semi-empirical DFT-D2 method from Grimme.[43] All Brillouin zone integrations were sampled using a (4×4×1) Γ-centered Monkhorst-Pack *k*-point scheme.[44]

Each QL was modeled as part of the R3m space group (no. 160). The bulk crystal was modeled using two formula units ($Z = 2$) with $\sigma_v$ symmetry [see Fig. 4(b)]; the layer inversion was required to remove spurious polarization effects in the slab supercells, thereby limiting *n* to even integers. The geometric optimization of the bulk crystal was performed until the Hellmann-Feynman forces converged to less than 0.02 eV/Å; the lattice constants were found to be $a$ = 3.987 Å and $c$ = 21.126 Å with an interlayer distance (Se-Se) of 3.65 Å. For the $n$ = 2, 4, 6, and 8 cases, further optimization did not noticeably alter the atomic positions; a vacuum spacing of 27 Å was included in the *z*-direction to minimize interactions with periodic images. The static dielectric constants and Born effective charge tensors were computed based on linear response theory under finite electric fields (= 0.005 eV/Å) which was implemented in VASP.[45]

## ASSOCIATED CONTENT

**Supporting Information**

The complete AFM and MIM images of $In_2Se_3$ nano-flakes, thickness fitting of different layers, detailed finite-element simulation, and the inhomogeneous MIM images of 4-5 layered $In_2Se_3$ are given in this section.

## AUTHOR INFORMATION

**Notes**

The authors declare no competing financial interest.

**Corresponding Author**

\*\*Email: gshwang@che.utexas.edu and kejilai@physics.utexas.edu

**Contributions**

*D. W. and A. P. contributed equally to this work.

ACKNOWLEDGEMENTS

The MIM work (DW, YL, XW, YR, KL) was supported by Welch Foundation Grant F-1814. Theoretical calculation (AP, YT, GH) was supported in part by the Robert A. Welch Foundation (F-1535). We would also like to thank the Texas Advanced Computing Center for use of their computing resources.

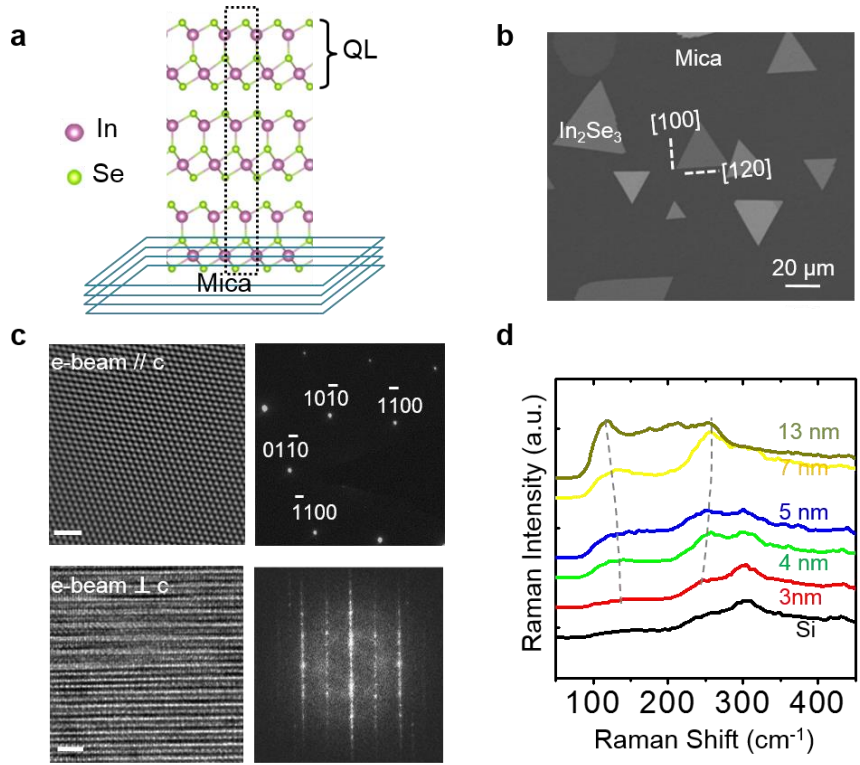

**Figure. 1** (a) Schematic of the few-layer In$_2$Se$_3$ sample on the mica substrate. Each quintuple layer (QL) comprises 5 atomic Se-In-Se-In-Se layers. The interlayer distance between layers is 2.87 Å in the z-direction. For the R3m lattice, the unit cell consists of three layers and is enclosed by the dashed box in the center. (b) SEM image of a typical discrete In$_2$Se$_3$ flakes on mica. (c) HRTEM images (left panel) and corresponding selected area electron diffraction (SAED, right-up) and fast Fourier transform (FFT, right-down) patterns of //c axis and ⊥c axis. (d) Raman spectra of In$_2$Se$_3$ flakes with different thicknesses transferred onto SiO$_2$ (300 nm)/Si substrates.

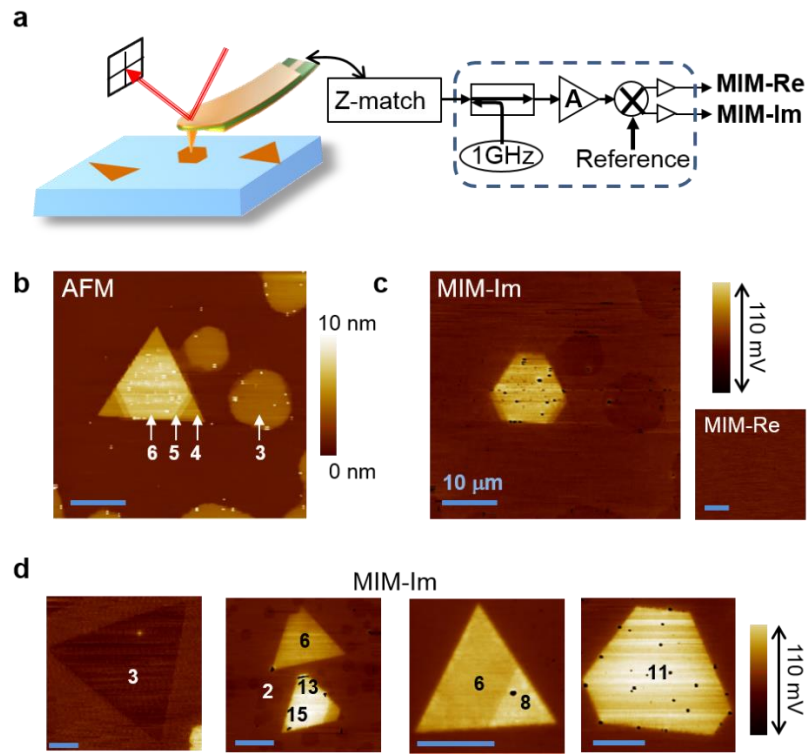

**Figure 2.** (a) Schematic of the simultaneous AFM and MIM measurements. (b) Topographic and (c) MIM-Im/Re images of the few-layer $In_2Se_3$ nano-flakes. The number ($n$) of layers is indicated in the AFM image. (d) More MIM-Im data on two batches of samples (see the text). $n$ is labeled for each terrace on the sample. All scale bars are 10 μm in the images.

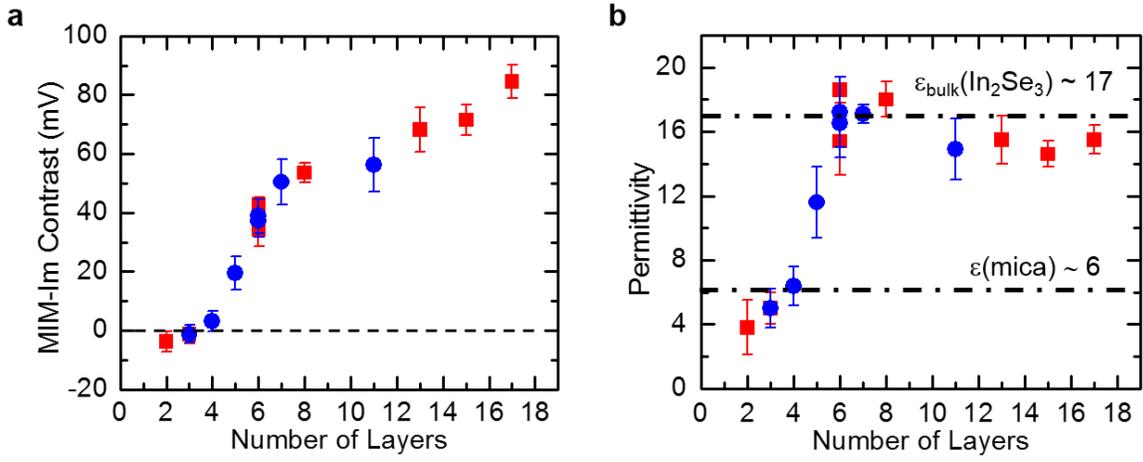

**Figure 3.** (a) MIM-Im contrast as a function of *n*. The dashed line is a guide to the eyes. (b) Permittivity deduced from MIM-Im as a function of *n*. The blue circles and red squares represent two different batches of samples.

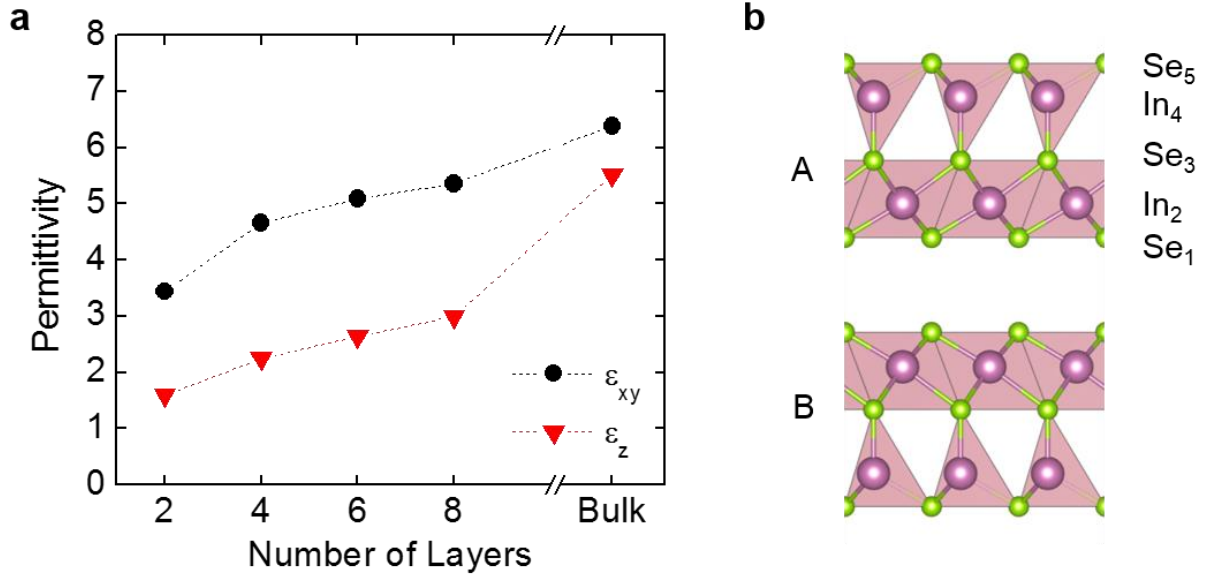

**Figure 4.** (a) Ion-clamped dielectric permittivity parallel ($\varepsilon_z$) and perpendicular ($\varepsilon_{xy}$) to the *c*-axis predicted from DFT-HSE06 calculations for varying number of QLs of $In_2Se_3$. (b) Representative schematic of the 2 QL case to demonstrate the use of inverted layering (i.e., A is a mirror of B). Purple and green balls represent In and Se atoms, respectively, while pink polyhedra indicate tetrahedral and octahedral sites. The numeric labels for Se and In are described in the manuscript.

**Table 1.** Born effective charges ($Z_\alpha^*$) for species within one QL of bilayer (n = 2) and bulk $In_2Se_3$.

|  | $Z_{xy}^*$ (n = 2) | $Z_{xy}^*$ (bulk) | $Z_z^*$ (n = 2) | $Z_z^*$ (bulk) |
|---|---|---|---|---|
| $Se_1$ | -2.54 | -2.54 | -0.25 | -0.66 |
| $In_2$ | 4.04 | 4.02 | 0.45 | 1.60 |
| $Se_3$ | -1.81 | -1.81 | -0.46 | -1.75 |
| $In_4$ | 2.81 | 2.79 | 0.57 | 1.79 |
| $Se_5$ | -2.50 | -2.49 | -0.30 | -0.94 |